\documentclass{psp-book9x6}
\usepackage{psp-book-van}
\usepackage{bm}

\title{New Challenges in Electrostatics of Soft and Disordered Matter}

\begin{document}


\setcounter{page}{1}

\chapter{The Wigner Strong-Coupling approach}

\chapauth{Ladislav \v{S}amaj$^{a,b}$ and Emmanuel Trizac$^{b}$
\chapaff{$^{a}$Institute of Physics, Slovak Academy of Sciences, Bratislava,
Slovakia\\ $^{b}$LPTMS, Universit\'e Paris-Sud, UMR CNRS 8626, Orsay, France}}

\section{Model}
Two equivalent charges in vacuum repeal each other. 
Let the two charges, say macro-ions, be immersed in an electrolyte of 
mobile micro-ions which is in thermal equilibrium at some inverse 
temperature $\beta = 1/(k_{\rm B}T)$.
Since the micro-ions are repelled/attracted by the macro-ions, 
the Coulomb interaction between macro-ions is modified by tracing out 
microscopic degrees of freedom. 
An important question is whether the ensuing effective interaction can become 
attractive in some distance range between macro-ions.
This question is essential in various fields of colloid science from
physics \cite{physics} to biochemistry \cite{biochemistry}. 
It was first answered from numerical investigations \cite{numerical} 
showing that the effective 
interaction, which is always repulsive in the high-temperature
(weak-coupling) region described adequately by the Poisson-Boltzmann 
mean-field theory \cite{Andelman06}, can become attractive  
at low temperatures (strong-coupling regime). 

\begin{figure}[tb]
\begin{center}
\psfig{file=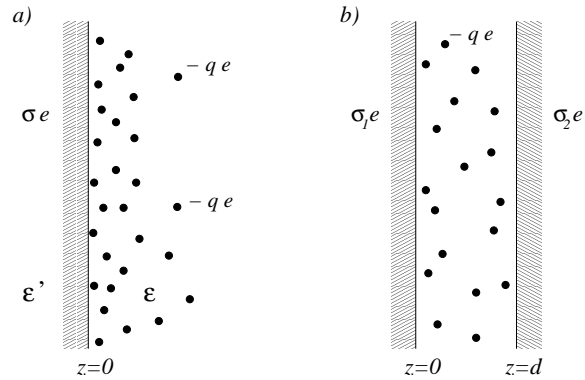,width=3in}
\end{center}
\caption{The two geometries considered: (a) one plate and (b) two parallel
plates at distance $d$.}
\label{figsamajtrizac.1}
\end{figure}  

A simplified model system may be proposed, for the sake of analytical 
understanding.
Firstly, spherical macro-ions are usually big colloids 
of several thousand elementary
charges $e$, and we represent their boundaries as plates with the
fixed surface charge density $\sigma e \simeq {\rm electron}/({\rm nm})^2$.
Secondly, we shall consider the no-salt regime, with 
only one type of mobile micro-ions (counterions) having 
charge $-q e$ (valence $q=1,2,\ldots$), 
under the requirement of global electroneutrality. 

The considered 3D geometries are pictured in Fig. \ref{figsamajtrizac.1}: 

\noindent $\bullet$
In the one-plate geometry (a), the charged $(x,y)$-surface 
is located at $z=0$.
The mobile counterions, confined to the half-space $z>0$, are immersed
in a solvent of dielectric permittivity $\epsilon$.
In realistic biological systems, the permittivity of the polarizable 
colloid $z<0$ ($\epsilon'\le 10$) is much smaller than that of the solvent
($\epsilon\simeq 80$ for water).

\noindent $\bullet$
In the geometry of two parallel plates (b), the counterions are
confined to the region $0<z<d$. 
Global electroneutrality requires that the surface charge densities of 
the plates are constrained by $\sigma_1+\sigma_2>0$.
Only the homogeneous dielectric case $\epsilon=\epsilon'$ will be worked out
for this geometry. 

Here, we shall concentrate on the strong-coupling (SC) regime,
defined as the limit where the appropriately defined coupling parameter
$\Xi$ [see Eq. (\ref{eq:Xidef})] is large.
Based on a field-theoretical representation, Netz and his collaborators 
\cite{VSC1} proposed a virial strong-coupling (VSC) approach.
A comparison with Monte Carlo (MC) simulations confirms the adequacy
of the leading SC behaviour which is a single-particle theory in 
the potential of the charged wall(s).
However, the VSC method fails to describe the subleading SC corrections.
The \emph{leading} order of the VSC theory was generalized to
asymmetrically charged plates, image charge effects, presence of salt
and curved (spherical and cylindrical) geometries \cite{VSC2}. 

Recently, we put forward a different SC approach \cite{WSC1,WSC2} based on 
a large coupling expansion in particle deviations around the ground state 
formed by the 2D Wigner crystal of counterions at the plate(s). It was hence
coined Wigner-SC (WSC). 
%
It will be shown that in doing so, one can address situations where the
VSC approach is invalid, such as the one plate problem with dielectric
mismatch. In cases where the VSC provides the correct dominant SC 
behaviour, the WSC yields the exact subsequent correction,
a task where VSC is unsuccessful. The reason is, in both cases
and even if the dominant SC behaviour stems from a single-particle picture, that the
problem should be envisioned as an $N$-body one, and does not comply by the
virial idea that 2-body interactions are a perturbative correction 
to one-body results. In this respect, the failure of the virial route
is here reminiscent of its breakdown for critical property studies 
of electrolytes \cite{FL}.

\section{One-plate geometry}

\subsection{Homogeneous dielectric case}
First we consider the one-plate geometry in Fig. \ref{figsamajtrizac.1} (a)
with no image charges, $\epsilon'=\epsilon$.
There are two relevant length scales describing counterion interactions. 
The potential energy of an isolated counterion at distance $z$ from the wall 
is given, in Gauss units, by the dimensionless relation
\begin{equation}
\beta E_1(z) = \frac{2\pi q \beta e^2\sigma}{\epsilon} z
\equiv \frac{z}{\mu} \equiv \widetilde{z} ,
\end{equation} 
where $\mu$ is the Gouy-Chapman length.
The interaction energy of two counterions at distance $r$ is given by
\begin{equation}
\beta E_2(r) = \beta \frac{(qe)^2}{\epsilon} \frac{1}{r}
\equiv \frac{q^2 \ell_{\rm B}}{r} , 
\end{equation}
where $\ell_{\rm B} = \beta e^2/\epsilon$ is the Bjerrum length.
This defines another length scale $q^2 \ell_{\rm B}$.
The dimensionless coupling parameter $\Xi$, quantifying the strength
of electrostatic correlations, is defined as the ratio
\begin{equation}
\Xi = \frac{q^2\ell_{\rm B}}{\mu} = 2\pi q^3 \ell_{\rm B}^2 \sigma .
\label{eq:Xidef}
\end{equation} 
The SC regime $\Xi\gg 1$ can be realized in a number of ways, 
e.g. low temperatures or large valence $q$/surface charge density $\sigma e$.

We are interested in the counterion density profile defined by
$\rho({\bf r}) \equiv \rho(z) = 
\langle \sum_{j=1}^N \delta({\bf r}-{\bf r}_j) \rangle$, where
$j$ runs over $N$ counterions at spatial positions $\{ {\bf r}_j \}$
and $\langle \cdots \rangle$ means thermal equilibrium average.
The profile will be considered in the rescaled form
$\widetilde{\rho}(\widetilde{z}) \equiv \rho(\mu \widetilde{z})/(2\pi\ell_{\rm B}\sigma^2)$.
The electroneutrality condition $q\int_0^{\infty} {\rm d}z\, \rho(z) = \sigma$
can be rewritten as follows
$\int_0^{\infty} {\rm d}\widetilde{z}\, \widetilde{\rho}(\widetilde{z}) = 1$.
The contact theorem \cite{contact} 
reads as $\beta P = \rho(0) - 2\pi\ell_{\rm B}\sigma^2$ and
since the fluid pressure $P$ vanishes for a single isolated 
double layer, we have $\widetilde{\rho}(0) = 1$.

\begin{figure}[tb]
\begin{center}
\psfig{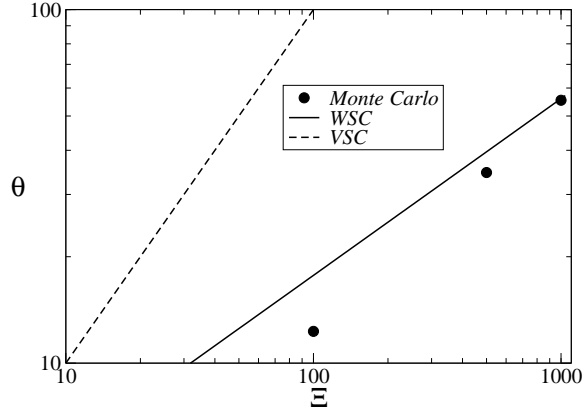}
\end{center}
\caption{The fitting parameter $\theta$ versus the coupling constant $\Xi$.}
\label{figsamajtrizac.2}
\end{figure}

According to the VSC method \cite{VSC1}, the density profile of counterions
can be formally expanded in the SC regime as a power series in $1/\Xi$:
\begin{equation} \label{VSCprofile}
\widetilde{\rho}(\widetilde{z},\Xi) = \widetilde{\rho}_0(\widetilde{z}) +
\frac{1}{\Xi} \widetilde{\rho}_1(\widetilde{z}) + O\left( \frac{1}{\Xi^2}\right) ,
\end{equation}
where
\begin{equation}
\widetilde{\rho}_0(\widetilde{z}) = {\rm e}^{-\widetilde{z}} , \quad
\widetilde{\rho}_1(\widetilde{z}) = {\rm e}^{-\widetilde{z}} 
\left( \frac{\widetilde{z}^2}{2} - \widetilde{z} \right) .
\end{equation}
While the leading single-particle term $\widetilde{\rho}_0(\widetilde{z})$ is in
agreement with MC simulations, the subleading $\widetilde{\rho}_1(\widetilde{z})$
has the expected functional form but the prefactor $1/\Xi$ is incorrect.
In particular, the MC data \cite{VSC1} were treated by using
$\widetilde{\rho}(\widetilde{z},\Xi)-\widetilde{\rho}_0(\widetilde{z}) = 
\widetilde{\rho}_1(\widetilde{z})/\theta$ with $\theta$ as a fitting parameter.
As is seen in the log-log plot of Fig. \ref{figsamajtrizac.2},
the MC numerical values of $\theta$ (filled circles) are much smaller
than the VSC prediction $\theta=\Xi$ (dashed line).
Note that the difference even grows with increasing $\Xi$.

The WSC approach \cite{WSC1} is based on the fact that in the asymptotic 
ground-state limit $\Xi\to\infty$ all counterions collapse on the oppositely 
charged surface $z=0$, forming a 2D hexagonal (equilateral triangular) Wigner
lattice \cite{Wigner}.
The lattice points are indexed by $j=(j_1,j_2)$, $j_1$ and $j_2$ being
any two integers: 
${\bf R}_j \equiv (R_j^x,R_j^y) = j_1 {\bm a}_1 + j_2 {\bm a}_2$, where 
${\bm a}_1= a(1,0)$ and ${\bm a}_2= (a/2)(1,\sqrt{3})$
are the primitive translation vectors of the Bravais lattice.
The lattice spacing $a$ is fixed by the condition of global electroneutrality,
$q = \sqrt{3} a^2 \sigma/2$.
Note that in the large-$\Xi$ limit, the lateral distance between 
nearest-neighbour counterions in the Wigner crystal $a$ is much larger
than the characteristic length $\mu$ in the perpendicular $z$-direction,
$a/\mu\propto \sqrt{\Xi}\gg 1$.

We denote the ground-state energy of counterions on the Wigner lattice
plus the fixed surface charge by $E_0$.
For $\Xi$ large but not infinite, vibrations of counterions around their 
lattice positions play a role.
Let us first shift one of the particles, say $j=1$, from its Wigner
lattice position $({\bf R}_1,z_1=0)$ by a small vector 
$\delta{\bf r}=(x,y,z>0)$ ($\vert \delta {\bf r}\vert \ll a$) and look for 
the corresponding change in the total energy $\delta E = E - E_0 \ge 0$.
The first contribution to $\delta E$ comes from the one-body interaction 
of the shifted counterions with the fixed surface charge density,
\begin{equation}
\delta E_1(z) = \frac{2\pi q e^2\sigma}{\epsilon} z .
\end{equation}
The second contribution comes from the two-body interactions of the
shifted counterions 1 with all other particles $j\ne 1$ on the 2D
hexagonal lattice:
\begin{eqnarray}
\delta E_2(x,y,z) & = & \frac{(qe)^2}{\epsilon} \sum_{j\ne 1} \left[ 
\frac{1}{\sqrt{(R_{1j}^x+x)^2+(R_{1j}^y+y)^2+z^2}} - \frac{1}{R_{1j}} \right]
\nonumber \\ & \sim & \frac{(qe)^2}{2\epsilon a^3} C_3 
\left[ \frac{1}{2}(x^2+y^2) - z^2 \right] ,
\end{eqnarray}
where ${\bf R}_{1j}\equiv {\bf R}_1-{\bf R}_j$ and the dimensionless
lattice sum $C_3 = \sum_{j\ne 1} (R_{1j}/a)^{-3}=11.034\ldots$.
It can be shown \cite{WSC1} that harmonic deviations in the $(x,y)$ plane
and higher-order deviation terms do not contribute to the first SC correction 
and we can write
\begin{equation}
-\beta \delta E = - \widetilde{z} + \frac{3^{3/4}C_3}{2(4\pi)^{3/2}}
\frac{1}{\sqrt{\Xi}} \widetilde{z}^2 + \cdots .
\end{equation}
The generalization to shifts of all $j=1,2,\ldots$ counterions from their 
lattice positions $({\bf R}_j,z_j=0)$ by a small vector 
$\delta {\bf r}_j=(x_j,y_j,z_j>0)$ ($\vert \delta {\bf r}_j\vert \ll a$) 
is straightforward and leads to
\begin{equation}
-\beta \delta E \sim - \sum_j \widetilde{z}_j  + \frac{3^{3/4}}{2(4\pi)^{3/2}}
\frac{1}{\sqrt{\Xi}} \sum_{j<k} 
\frac{(\widetilde{z}_j-\widetilde{z}_k)^2}{(\vert {\bf R}_j-{\bf R}_k \vert/a)^3} 
+ \cdots .
\end{equation}
Two-body interactions are only a perturbation with respect to the one-body 
terms which explains why a single-particle picture provides the leading SC 
behaviour.
Introducing a generating field \cite{WSC1}, the cumulant perturbation
expansion yields the density profile 
\begin{equation} \label{WSCprofile}
\widetilde{\rho}(\widetilde{z}) =
\underbrace{{\rm e}^{-\widetilde{z}}}_{\widetilde{\rho}_0(\widetilde{z})} +
\underbrace{\frac{3^{3/4}}{(4\pi)^{3/2}} \frac{C_3}{\sqrt{\Xi}}}_{1/\theta}
\underbrace{{\rm e}^{-\widetilde{z}}\left( \frac{\widetilde{z}^2}{2}-\widetilde{z} 
\right)}_{\widetilde{\rho}_1(\widetilde{z})} + \cdots .
\end{equation}
Comparing with the VSC result (\ref{VSCprofile}) we see that the first
corrections have the same functional dependence on $\widetilde{z}$, but
different prefactors.
In terms of the fitting parameter $\theta$, the VSC estimate $\theta=\Xi$
is compared with the present value
\begin{equation}
\theta = \frac{(4\pi)^{3/2}}{3^{3/4}} \frac{1}{C_3} \sqrt{\Xi}
= 1.771\ldots \sqrt{\Xi} .
\end{equation}
As is seen from Fig. \ref{figsamajtrizac.2}, this formula (solid curve)
is in full agreement with MC data (filled circles).

\subsection{Dielectric inhomogeneity}
In the inhomogeneous dielectric case, we define the dielectric jump
between the solvent and the wall as
$\Delta = (\epsilon-\epsilon')/(\epsilon+\epsilon')$.
The image of a charge $e$ at position ${\bf r}=(x,y,z>0)$ has
the charge of strength $e^*=e\Delta$ and is localized at 
${\bf r}^*=(x,y,-z<0)$.
Counterions interact via the Coulomb interaction potential
$u({\bf r},{\bf r}') = u_0({\bf r},{\bf r}') + u_{\rm im}({\bf r},{\bf r}')$,
where 
\begin{equation}
u_0({\bf r},{\bf r}') = \frac{1}{\varepsilon\vert{\bf r}-{\bf r}'\vert} , 
\quad u_{\rm im}({\bf r},{\bf r}') = 
\frac{\Delta}{\varepsilon\vert{\bf r}^*-{\bf r}'\vert}
\end{equation}
are the direct and image Coulomb potentials, respectively. 
The total interaction energy of the counterions at positions 
$\{ {\bf r}_j\}_{j=1}^N$ in the half-space $z>0$ reads
\begin{equation} \label{eq:energypi}
E =  \sum_{j=1}^N \left[ \frac{2\pi qe^2\sigma}{\varepsilon} (1+\Delta) z_j
+ \frac{(qe)^2}{4\varepsilon} \frac{\Delta}{z_j} \right] 
+ \frac{(qe)^2}{2} \sum_{\genfrac{}{}{0pt}{}{j,k=1}{(j\ne k)}}^N 
u({\bf r}_j,{\bf r}_k) .
\end{equation}
Here, the two one-body terms in the square brackets originate in 
the interaction of the particle with the surface charge plus its image 
(therefore the factor $1+\Delta$) and with its own charge image.

We shall restrict ourselves to the case of repulsive $\Delta>0$ 
image charges which is of special interest.
The counterions are on the one hand attracted to the wall by 
the fixed surface charge $\sigma e$ and on the other hand repelled from 
the wall by their images.
It is natural to assume that the ground state of the system is formed by
the standard 2D hexagonal Wigner crystal of counterions with lattice spacing 
$a$, localized at some nonzero distance $l$ from the wall.
This distance is determined by balancing the attractive and repulsive forces.
The energy per particle for the Wigner crystal at distance $z$ from the wall 
is given by
\begin{eqnarray}
\frac{\beta E_0(z)}{N} & = & (1+\Delta)\frac{z}{\mu} 
+ \frac{q^2 \ell_{\rm B}}{2} \sum_{j\ne 1} \frac{1}{R_{1j}} \nonumber \\ 
& & + \frac{q^2 \ell_{\rm B}}{2} \Delta \left[ \frac{1}{2z} + 
\sum_{j\ne 1} \frac{1}{\sqrt{R_{1j}^2+(2z)^2}} \right] .
\end{eqnarray} 
On the rhs, the first term describes the interaction of the
particle with the surface charge, the second ($z$-independent) one
the direct interaction with all other particles, and the last two terms 
the interaction of the particle with its own self-image and with
images of all other particles. 
The distance of the Wigner crystal from the wall $l$ is determined by 
the stationarity condition $\partial_z [\beta E_0(z)/N]\vert_{z=l}=0$.
Introducing the variable $t=2l/a$, this requirement can be written as
\begin{equation} \label{eq:deft}
\sum_{j,k=-\infty}^{\infty} \frac{1}{[t^2+(j^2+jk+k^2)]^{3/2}}
= \frac{4\pi}{\sqrt{3}} \frac{1+\Delta}{\Delta}\frac{1}{t} ,
\end{equation}
where use was made of the hexagonal structure.
The lattice sum appearing here can be represented as an integral over 
Jacobi theta functions \cite{WSC2}, which can be 
evaluated with a high precision. 
The numerical solution of Eq. (\ref {eq:deft}) for the extreme case 
$\Delta=1$ is $t\simeq 0.295$.
In the limit of small $\Delta$, we have $t\sim 3^{1/4}\sqrt{\Delta/(4\pi)}$.
To examine the ground-state stability of the Wigner crystal along 
the $z$ direction, we shift one particle (say $j=1$) from its lattice 
position by a small deviation $z-l$.
The corresponding change in the energy $\beta\delta E = (z-l)^2/\xi^2$
turns out to be positive. The Wigner lattice is therefore
thermodynamically stable, 
for all $\Delta\in [0,1]$ \cite{WSC2}. 

\begin{figure}[tb]
\begin{center}
\psfig{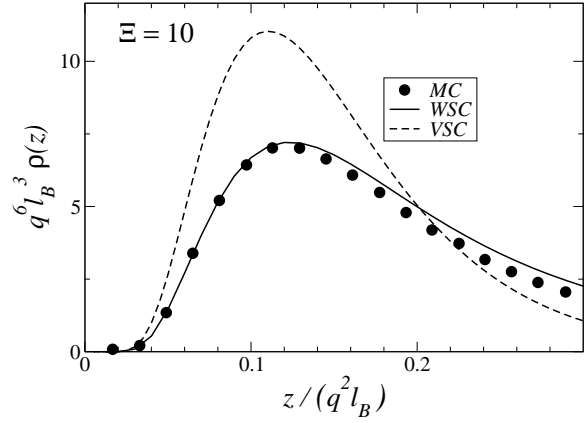}
\end{center}
\caption{The density profile in various approaches, for the dielectric 
jump $\Delta=0.95$ and the coupling constant $\Xi=10$. The value $q=1$ 
was used in \protect\cite{VSC3} for the Monte Carlo (MC) simulations.}
\label{figsamajtrizac.3}
\end{figure}

The distance of the Wigner crystal from the wall $l$ is of the order of 
the Wigner lattice spacing $a$.
Shifting a particle from its lattice position along the $z$-axis, 
its pair interactions with all counterions is of the same relevance as 
its one-body interaction with the charged wall. We thus
anticipate the failure of the VSC route.
In the leading WSC order, we consider the particle potential induced by the 
charged wall plus the frozen hexagonal Wigner crystal (excluding the
particle under consideration) at distance $l$ from the wall:
\begin{eqnarray} \label{potential}
\beta u(z) & = & (1+\Delta) \frac{z}{\mu} + q^2 \ell_{\rm B} 
\sum_{j\ne 1} \frac{1}{\sqrt{R_{1j}^2+(z-l)^2}} \nonumber \\
& & + q^2 \ell_{\rm B} \Delta \left[ \frac{1}{4 z} +
\sum_{j\ne 1} \frac{1}{\sqrt{R_{1j}^2+(z+l)^2}} \right] .
\end{eqnarray}
We have at our disposal MC results of the density profile only for the dielectric 
jump $\Delta=0.95$ and the relatively small coupling $\Xi=10$ \cite{VSC3}, 
see filled circles in Fig. \ref{figsamajtrizac.3}.
The VSC density profile is represented by the dashed curve,
in disagreement with MC. On the other hand,
the WSC density profile (solid curve) is calculated by using 
the potential (\ref{potential})
and shows good agreement with MC data up to
$z\sim 0.2 q^2\ell_{\rm B} = 2\mu$, even for such a small value of $\Xi$.
%

\section{Two-plate geometry}
We next consider symmetric like-charged plates 
$\sigma_1=\sigma_2\equiv \sigma$.
The electric field between the plates vanishes. 
At $T=0$, the classical system is defined furthermore by
the dimensionless separation
\begin{equation}
\eta = d \sqrt{\frac{\sigma}{q}} = \frac{1}{\sqrt{2\pi}}
\frac{\widetilde{d}}{\sqrt{\Xi}} .
\end{equation}
A minor complication comes from the fact that counterions form, on
the opposite plate surfaces, a bilayer Wigner crystal whose 
structure depends on $\eta$ \cite{bilayer}.
Here, we aim at performing expansions of thermodynamic quantities in powers 
of $\eta\propto \widetilde{d}/\sqrt{\Xi}\ll 1$ since 
we fix the scale $\widetilde{d}$ 
while $\Xi$ becomes large.
At the smallest separation $\eta=0$, a single hexagonal Wigner crystal
is formed.
Its lattice spacing $b$ is determined by global neutrality as
$q = \sqrt{3} b^2 \sigma$.
Since $\eta\ll 1$ is equivalent to $d/b\ll 1$, we shift particles along 
the $z$-axis around this structure. 
The corresponding energy change is
\begin{equation}
\delta E = \mbox{cst.} - \frac{(q e)^2}{4\epsilon} 
\sum_{\genfrac{}{}{0pt}{}{j,k=1}{j\ne k}}^N \frac{(z_j-z_k)^2}{
\vert {\bf R}_j-{\bf R}_k\vert^3} .
\end{equation}
The cumulant technique \cite{WSC1} then implies the density profile
\begin{equation}
\widetilde{\rho}(\widetilde{z}) = \frac{2}{\widetilde{d}} + \frac{1}{\theta}
\frac{2}{\widetilde{d}} \left[ \left( \widetilde{z}-\frac{\widetilde{d}}{2} \right)^2
- \frac{\widetilde{d}^2}{12} \right] + O\left( \frac{\widetilde d^2}{\theta^{\,3/2}} \right) ,
\end{equation}
where the $\theta$-parameter is now obtained in the form
\begin{equation}
\theta = \frac{(4\pi)^{3/2}}{3^{3/4}} \frac{1}{C_3} \frac{1}{\sqrt{2}} 
\sqrt{\Xi} = 1.252\ldots \sqrt{\Xi} .
\end{equation}
The previous VSC result was $\theta=\Xi$ \cite{VSC1}.
Defining the rescaled (temperature-independent) pressure
$\widetilde{P} \equiv \beta P/(2\pi\ell_{\rm B}\sigma^2)$, the contact-value theorem 
$\widetilde{P}=\widetilde{\rho}(0)-1=\widetilde{\rho}(\widetilde{d})-1$ implies
\begin{equation}
\widetilde{P} = -1 + \frac{2}{\widetilde{d}} + \frac{\widetilde{d}}{3\theta}
+ O\left( \frac{\widetilde d^2}{\theta^{\,3/2}} \right) .
\label{eq:eos}
\end{equation}

\begin{figure}[tb]
\begin{center}
\psfig{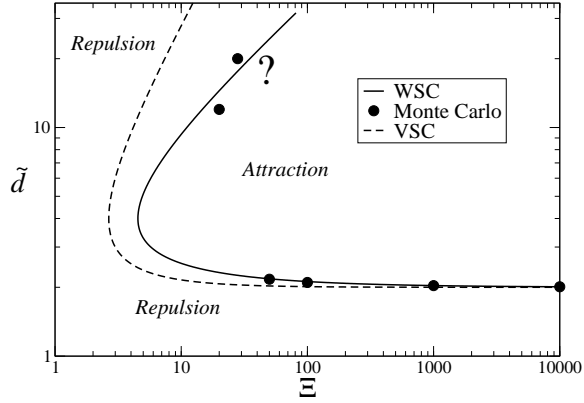}
\end{center}
\caption{Phase diagram in the $(\Xi,\widetilde{d})$ plane for symmetric 
like-charged plates.}
\label{figsamajtrizac.4}
\end{figure}  

The phase diagram in the $(\Xi,\widetilde{d})$ plane, following from this
WSC equation, is shown in Fig. \ref{figsamajtrizac.4}.
The shape of the isobaric phase boundary $P=0$ (solid curve), which divides
the plane onto its attractive $(P<0)$ and repulsive $(P>0)$ parts,
shows striking similarity with its counterpart obtained 
numerically \cite{VSC1} (filled circles).
The question mark is a reminder that the upper branch 
where $\widetilde{d}\propto\sqrt{\Xi}$ exceeds the validity domain 
of Eq. (\ref{eq:eos}), which is $\widetilde{d}\propto\Xi^{1/4}$.
The dashed line is the inaccurate VSC prediction.

The WSC formalism can be readily generalized to plates 
with the asymmetry parameter $\zeta=\sigma_2/\sigma_1 \in [-1,1]$.
The final result for the rescaled pressure reads
$\widetilde{P} = \widetilde{P}_0 + \widetilde{P}_1/\sqrt{\Xi} + O(1/\Xi)$, where
\begin{equation} \label{eq:p0}
\widetilde{P}_0 = - \frac{1}{2} (1+\zeta^2) + 
\frac{1}{2} (1-\zeta^2) \coth\left( \frac{1-\zeta}{2}\widetilde{d} \right)
\end{equation}
is the leading SC contribution, already obtained within the VSC
method \cite{VSC2}, and the coefficient of the first $1/\sqrt{\Xi}$ correction
\begin{eqnarray}
\widetilde{P}_1 & = & \frac{3^{3/4} (1+\zeta)^{5/2} C_3}{4 (4\pi)^{3/2}} 
\frac{\widetilde{d}}{\sinh^2\left(\frac{1-\zeta}{2}\widetilde{d} \right)} 
\nonumber \\ & & \times 
\left[ \left( \frac{1-\zeta}{2}\widetilde{d} \right) 
\coth\left( \frac{1-\zeta}{2}\widetilde{d} \right) - 1 \right] . \label{eq:p1}
\end{eqnarray}

\begin{figure}[tb]
\begin{center}
\psfig{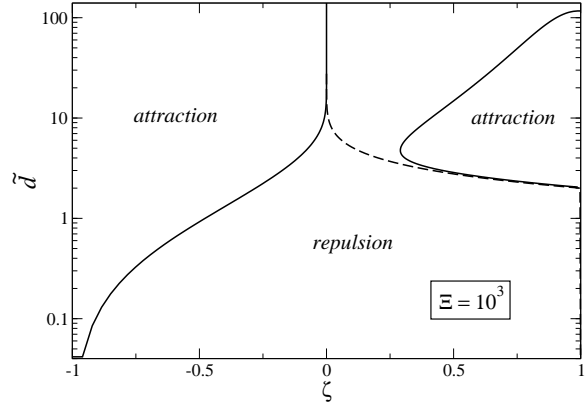}
\end{center}
\caption{Phase diagram of asymmetrically charged plates. 
}
\label{figsamajtrizac.5}
\end{figure}  

For the coupling constant $\Xi=10^3$, the phase diagram in the whole
range of the asymmetry parameter $\zeta$ is shown in
Fig. \ref{figsamajtrizac.5}.
The dashed line corresponds to the leading-order (common to VSC and WSC)
result of the phase boundary $\widetilde{P}_0=0$, which is equivalent to
$\widetilde{d}^* = - 2 \ln\vert\zeta\vert/(1-\zeta)$. 
The solid line corresponds to the full WSC phase boundary 
$\widetilde{P}=\widetilde{P}_0+\widetilde{P}_1/\sqrt{\Xi}=0$.
For oppositely charged plates $-1<\zeta\le 0$, the difference between
the solid and dashed curves is invisible.
On the other hand, the first correction affects significantly
the positive $\zeta$ part of the phase diagram: 
no like-charge attraction is found in the range $0<\zeta<0.29$, 
whereas the leading-order result (dashed line) predicts 
an attraction region for all $\zeta$. 

\section{Conclusion}
Although the Wigner crystal becomes thermodynamically unstable for
relatively large coupling constants $\Xi\sim 10^5-10^6$, it can serve
as the starting point for WSC expansions valid for much smaller values
of $\Xi\sim 10-100$.
This is remarkable but 
not exceptional in statistical mechanics, e.g. a low-temperature 
expansion of an Ising model around its ground state can be interpolated into
the critical region.

Here are the main advantages of our WSC approach comparing to 
the original VSC method.

\noindent $\bullet$ The WSC approach is technically simple, the treatment of harmonic 
and higher vibrations requires an elementary technique of cumulants.
On the other hand, there were no attempts to go beyond the leading term in nontrivial 
applications of the technically complicated VSC method such as for asymmetric plates.

\noindent $\bullet$ In contrast to the VSC method, the WSC theory implies accurately also
the first correction to the leading SC behavior, 
see Fig. \ref{figsamajtrizac.2}.
As was shown for two asymmetric like-charged plates in 
Fig. \ref{figsamajtrizac.5}, this correction can significantly modify 
the phase diagram. 

\noindent $\bullet$ More importantly, the WSC method applies also to physical situations when the leading SC 
description is not of single-particle type.
We documented this on the one-plate geometry with repulsive image charges,
where the Wigner structure formed at a finite distance from the plate 
contributes to the leading order (Fig. \ref{figsamajtrizac.3}).
There, the VSC approach fails severely.
Other problems of this kind include curved surfaces,
and will be the focus of future work.

\end{document}